# A SIMO Fiber Aided Wireless Network Architecture

Siddharth Ray, Muriel Médard and Lizhong Zheng
Laboratory for Information and Decision Systems
Massachusetts Institute of Technology, Cambridge, MA 02139.
Email: sray@mit.edu, medard@mit.edu, lizhong@mit.edu

*Abstract*— The concept of a fiber aided wireless network architecture (FAWNA) is introduced in [Ray et al., *Allerton Conference 2005*], which allows high-speed mobile connectivity by leveraging the speed of optical networks. In this paper, we consider a single-input, multiple-output (SIMO) FAWNA, which consists of a SIMO wireless channel and an optical fiber channel, connected through wireless-optical interfaces. We propose a scheme where the received wireless signal at each interface is quantized and sent over the fiber. Though our architecture is similar to that of the classical CEO problem, our problem is different from it. We show that the capacity of our scheme approaches the capacity of the architecture, exponentially with fiber capacity. We also show that for a given fiber capacity, there is an optimal operating wireless bandwidth and an optimal number of wireless-optical interfaces. The wireless-optical interfaces of our scheme have low complexity and do not require knowledge of the transmitter code book. They are also extendable to FAWNAs with large number of transmitters and interfaces and, offer adaptability to variable rates, changing channel conditions and node positions.

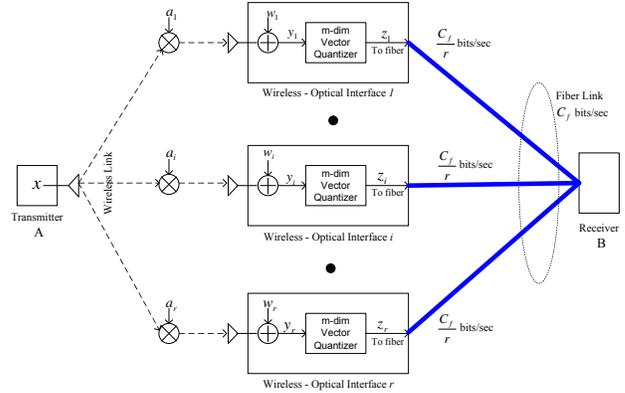

Fig. 1. A SIMO fiber aided wireless network architecture.

## I. INTRODUCTION

There is a considerable demand for increasingly high-speed communication networks with mobile connectivity. Traditionally, high-speed communication has been efficiently provided through wireline infrastructure, particularly based on optical fiber, where bandwidth is plentiful and inexpensive. However, such infrastructure does not support mobility. Instead, mobile communication is provided by wireless infrastructure, most typically over the radio spectrum. However, limited available spectrum and interference effects limit mobile communication to lower data rates.

We introduce the concept of a fiber aided wireless network architecture (FAWNA) in [10], which allows high-speed mobile connectivity by leveraging the speed of optical networks. Optical networks have speeds typically in hundreds of Megabit/sec or several Gigabit/sec (Gigabit Ethernet, OC-48, OC-192, etc.). In the proposed architecture, the network coverage area is divided into zones such that an optical fiber "bus" passes through each zone. Connected to the end of the fiber is a bus controller/processor, which coordinates use of the fiber as well as connectivity to the outside world. Along the fiber are radio-optical converters (wireless-optical interfaces), which are access points consisting of simple antennas directly connected to the fiber. Each of these antennas harvest the energy from the wireless domain to acquire the full radio bandwidth in their local environment and place the associated waveform onto a subchannel of the fiber. Within the fiber, the harvested signals can be manipulated by the bus controller/processor and made available to all other antennas. In each zone, there may be one or more active wireless nodes. Wireless nodes communicate between one another, or to the outside world, by communicating to a nearby antenna. Thus any node in the network is at most two hops away from any other node, regardless of the size of the network. In general, each zone is generally covered by several antennas, and there may also be wired nodes connected directly to the fiber.

This architecture has the potential to reduce dramatically the interference effects that limit scalability and the energy-consumption characteristics that limit battery life, in pure wireless infrastructure. A FAWNA uses the wireline infrastructure to provide a distributed means of aggressively harvesting energy from the wireless medium in areas where there is a rich, highly vascularized wireline infrastructure and distributing in an effective manner energy to the wireless domain by making use of the proximity of transmitters to reduce interference.

In this paper, we consider a single-input, multiple-output (SIMO) fiber aided wireless network architecture. We will also refer to this as SIMO-FAWNA. Figure 1 shows such a link between two points A and B. The various quantities in the figure will be described in detail in the next section. In the two hop link, the first hop is over a wireless channel and the second, over a fiber optic channel. The links we consider are ones where the fiber optic channel capacity is larger than the wireless channel capacity.

The transmitter at A transmits information to intermediate wireless-optical interfaces over a wireless SIMO channel. The wireless-optical interfaces then relay this information to the destination, B, over a fiber optic channel. The end-to-end design is done to maximize the transmission rate between A and B. Since a FAWNA has a large number of wireless-optical interfaces, an important design objective is to keep the wireless-optical interface as simple as possible without sacrificing too much in performance.

Our problem has a similar setup, but a different objective than the CEO problem [9]. In the CEO problem, the rate-distortion tradeoff is analyzed for a given source that needs to be conveyed to the CEO through an asymptotically large number of agents. Rate-distortion theory, which uses infinite dimensional vector quantization, is used to analyze the problem. We instead compute the maximum end-to-end rate at which reliable communication is possible. In general, duality between the two problems doesn't exist. Unlike the CEO problem, the number of wireless-optical interfaces is finite and the rate (from interface to receiver B) per interface is high due to the fiber capacity being large. Finite-dimensional, high resolution quantizers are used at the interfaces.

Let us denote the capacities of the wireless and optical channels as $C_w(P,W,r)$ and $C_f$ bits/sec, respectively, where, $P$ is the average transmit power at A, $W$ is the wireless transmission bandwidth and $r$ is the number of wireless-optical interfaces. Since, as stated earlier, we consider links where $C_w(P,W,r) \leq C_f$, the capacity of a SIMO-FAWNA, $C_{\mathsf{SIMO}}(P,W,r,C_f)$, can be upper bounded as

$$C_{\mathsf{SIMO}}(P,W,r,C_f) < \min\left\{C_w(P,W,r), C_f\right\} = C_w(P,W,r) \text{ bits/sec.} \quad (1)$$

One way of communicating over a SIMO-FAWNA is to decode and re-encode at the wireless-optical interface. A major drawback of the decode/re-encode scheme is significant loss in optimality because "soft" information in the wireless signal is completely lost by decoding at the wireless-optical interface. Hence, multiple antenna gain is lost. Moreover, decoding results in the wireless-optical interface having high complexity and the interface requires knowledge of the transmitter code book.

In this paper, we propose a scheme where the wireless signal at each wireless-optical interface is sampled and quantized using a fixed-rate, memoryless, vector quantizer, before being sent over the fiber. Hence, the interfaces use a forwarding scheme. Since transmission of continuous values over the fiber is practically not possible using commercial lasers, quantization is necessary for the implementation of a forwarding scheme in a FAWNA. The proposed scheme thus has quantization between the end-to-end coding and decoding. Knowledge of the transmitter code book is not required at the wireless-optical interface. The loss in "soft" information due to quantization of the wireless signal, goes to 0 asymptotically with increase in fiber capacity. The interface has low complexity, is practically implementable, is extendable to FAWNAs with large number of transmitters and interfaces and, offers adaptability to variable rates, changing channel conditions and node positions.

We show that the capacity using our scheme approaches the upper bound (1), exponentially with fiber capacity. The proposed scheme is thus near-optimal since, the fiber capacity is larger than the wireless capacity. Low dimensional (or even scalar) quantization can be done at the interfaces without significant loss in performance. Not only does this result in low complexity, but also smaller (or no) buffers are required, thereby further simplifying the interface. Hui and Neuhoff [8] show that asymptotically optimal quantization can be implemented with complexity increasing at most polynomially with the rate. For a SIMO-FAWNA with fixed fiber capacity, quantizer distortion as well as wireless capacity, $C_w(P,W,r)$, increases with wireless bandwidth and number of interfaces. The two competing effects result in the existence of an optimal operating wireless bandwidth and an optimal number of wireless-optical interfaces.

The paper is organized as follows: In section II, we describe our wireless and fiber channel models. We describe our scheme in section III and analyze its performance in section IV. We conclude in section V. Unless specified otherwise, all logarithms in this paper are to the base 2.

## II. CHANNEL MODEL

There are $r$ wireless-optical interfaces and each one of them is equipped with a single antenna. The interfaces relay the wireless signals they receive from the transmitter, to receiver B, over an optical fiber. Communication over the fiber is interference free, which may be achieved, for example, using Time Division Multiple Access (TDMA) or Frequency Division Multiple Access (FDMA).

*Wireless Channel:* We use a linear model for the wireless channel between A and the wireless-optical interfaces:

$$\vec{y} = \vec{a}\mathbf{x} + \vec{w}, \quad (2)$$

where, $\mathbf{x} \in \mathcal{C}, \vec{w}, \vec{y} \in \mathcal{C}^r$ are the channel input, noise and output, respectively. The channel gain vector (state), $\vec{a} \in \mathcal{C}^r$, is fixed and perfectly known at the receiver. The noise, $\vec{w}$, is a zero mean complex Gaussian random vector, $\vec{w} \sim \mathcal{CN}(0, N_0 I_r)$, and is independent of the channel input. $N_0/2$ is the double-sided white noise spectral density and $I_r$ is a $r \times r$ identity matrix. The channel input, $\mathbf{x}$, satisfies the average power constraint $E[|\mathbf{x}|^2] = P/W$, where, $P$ and $W$ are the average transmit power at A and wireless bandwidth, respectively. Hence, the wireless channel capacity is

$$C_w(P,W,r) = W \log\left(1 + \frac{\|\vec{a}\|^2 P}{N_0 W}\right),$$

and $W$ symbols/sec are transmitted over the wireless channel. Thus, using (1), we obtain an upper bound to the SIMO-FAWNA capacity:

$$C_{\mathsf{SIMO}}(P,W,r,C_f) < W \log\left(1 + \frac{\|\vec{a}\|^2 P}{N_0 W}\right). \quad (3)$$

*Fiber Optic Channel:* The fiber optic channel between the wireless-optical interface and the receiver, B, can reliably support a rate of $C_f$ bits/sec. Communication over the fiber is interference free and, the fiber capacity is equally divided among the interfaces. To keep the interfaces simple, source coding is not done at the interfaces. Later we show that since the fiber capacity is much larger compared to the wireless capacity, the loss from no source coding is negligible. Hence, at most $C_f/r$ bits/sec can be reliably communicated from any wireless-optical interface to the receiver, B. Fiber channel coding is performed at the wireless-optical interface to achieve this. Note that the code required for the fiber is a very low complexity one. An example of a code that may be used is the 8B10B code, which is commonly used in Ethernet. Hence, fiber channel coding does not significant increase the complexity at the wireless-optical interface.

## III. PROPOSED SCHEME

The input to the wireless channel, $\mathbf{x}$, is a zero mean circularly symmetric complex Gaussian random variable, $\mathbf{x} \sim \mathcal{CN}(0, P/W)$. Note that it is this input distribution that achieves the capacity of our wireless channel model.

All wireless-optical interfaces have identical construction. At each wireless-optical interface, the output from the antenna is first converted from passband to baseband and then sampled at the Nyquist rate of $W$ complex samples/sec. The random variable, $\mathbf{y}_i$, represents the output from the sampler at the $i^{th}$ interface. Fixed-rate, memoryless, $m$-dimensional vector quantization is performed on these samples at a rate of $l$ bits/complex sample. The quantized complex samples are subsequently sent over the fiber after fiber channel coding and modulation.

Hence, the fiber is required to reliably support a rate of $Wl$ bits/sec from each wireless-optical interface to the receiver, B. Since, identical quantizers are used at the wireless-optical interfaces, we get the following constraint on $l$:

$$l \leq \frac{C_f}{rW}. \quad (4)$$

The quantizer noise at the $i^{th}$ interface, $\mathbf{q}_{i,m,l}$, is modelled as being additive. Hence, the two-hop channel between A and B can be modelled as:

$$\vec{z} = \vec{a}\mathbf{x} + \vec{w} + \vec{q}_{m,l}, \quad (5)$$

where, $\vec{q}_{m,l} = [\mathbf{q}_{1,m,l}, \ldots, \mathbf{q}_{r,m,l}]^T$, and $^T$ denotes transpose. The quantizer used at the interface is an optimal fixed rate, memoryless, $m$-dimensional, high resolution vector quantizer. Hence, its distortion-rate function is given by the Zador-Gersho function [1], [3], [5]:

$$E[|\mathbf{q}_{i,m,l}|^2] = E[|\mathbf{y}_i|^2] M_m \beta_m 2^{-l}$$
$$= \left(N_0 + \frac{|a_i|^2 P}{W}\right) M_m \beta_m 2^{-l}, \ i \in \{1, \ldots, r\} \quad (6)$$

$M_m$ is Gersho's constant, which is independent of the distribution of $\mathbf{y}_i$. $\beta_m$ is the Zador's factor that depends on the distribution of $\mathbf{y}_i$. Since the fiber channel capacity is large, the assumption that the quantizer is a high resolution one, is valid. Hence, $l \gg 1$ and $C_f \gg rW$. Since this quantizer is an optimal fixed rate, memoryless vector quantizer, references [2], [3], [4], [6], [7] show that it has the following properties, $E[\mathbf{q}_{i,m,l}] = 0$, $E[\mathbf{z}_i \mathbf{q}^*_{i,m,l}] = 0$ and $E[\mathbf{y}_i \mathbf{q}^*_{i,m,l}] = -E[|\mathbf{q}_{i,m,l}|^2]$. Hence, $E[|\mathbf{z}_i|^2] = E[|\mathbf{y}_i|^2] - E[|\mathbf{q}_{i,m,l}|^2]$.

Let the capacity for this scheme (in bits/sec) be denoted as $C_Q(P, W, r, m, l)$, where, $m \in \{1, 2, \ldots\}$ and $l \leq \frac{C_f}{rW}$. The following notation will be used for representing a diagonal matrix: $\text{Diag}\{v_1, \ldots, v_n\}$ is a $n \times n$ diagonal matrix with $v_1, \ldots, v_n$ as its diagonal elements. We establish the following theorem. The proof is omitted for brevity.

*Theorem 1:*

$$C_{\text{SIMO}}(P, W, r, C_f) \geq C_Q(P, W, r, m, l)$$
$$\geq W \log\left(1 + \frac{\|\vec{a}\|^2 P}{N_0 W}\right) - \Phi(P, W, r, m, l), (7)$$

where,

$$\Phi(P, W, r, m, l) = W \log\left(1 + \frac{\|\vec{a}\|^2 P}{N_0 W}\right)$$
$$+ W \log\left(1 - \frac{P(1 - M_m \beta_m 2^{-l})}{N_0 W} \vec{a}^\dagger \left[\frac{P(1 - M_m \beta_m 2^{-l})}{N_0 W} \vec{a}\vec{a}^\dagger \right. \right.$$
$$\left. \left. + I_r + \frac{P M_m \beta_m 2^{-l}}{N_0 W} \text{Diag}\{|a_1|^2, \ldots, |a_r|^2\}\right]^{-1} \vec{a}\right). \square$$

## IV. PERFORMANCE ANALYSIS

In this section, we analyze the performance of the scheme described in Theorem 1. We examine how the capacity lower bound (7) is influenced by the quantization dimension and rate, transmit power, number of interfaces and wireless bandwidth. To simplify analysis, we will set the wireless channel gain, $\vec{a} = \vec{1}$, where, $\vec{1}$ is the $r$ dimensional column vector with all ones. Hence,

$$\Phi(P, W, r, m, l) = W \log\left(1 + \frac{rP}{N_0 W}\right) \quad (8)$$
$$- W \log\left(1 + \frac{r(1 - M_m \beta_m 2^{-l})\frac{P}{N_0 W}}{1 + \frac{P M_m \beta_m 2^{-l}}{N_0 W}}\right).$$

### A. Effect of quantizer dimension

To study the effect of quantizer dimension, $m$, on the performance of the proposed scheme, it suffices to consider the function, $\Phi(P, W, r, m, l)$. Since Gaussian signaling is used for the wireless channel, the input to the quantizer at the interface is a correlated Gaussian random vector. Zador's factor and Gersho's constant obey the following property: $M_\infty \beta_\infty \leq M_m \beta_m \leq M_1 \beta_1 \leq M_1 \beta_1^G$, where, $\beta_1^G$ is the Zador's factor for an i.i.d Gaussian source and, $\beta_1 \leq \beta_1^G$. $M_m \beta_m$ decreases with increase in $m$. Since, $M_1 = \frac{1}{12}$, $M_\infty = \frac{1}{2\pi e}$, $\beta_1^G = 6\sqrt{3}\pi$ and $\beta_\infty = 2\pi e$, $1 \leq M_m \beta_m \leq \frac{\pi\sqrt{3}}{2}$. The lower bound corresponds to fixed rate, infinite dimensional vector quantization, whereas, the upper bound corresponds to fixed rate, scalar quantization. In (8), $\frac{r(1 - M_m \beta_m 2^{-l})\frac{P}{N_0 W}}{1 + \frac{P M_m \beta_m 2^{-l}}{N_0 W}}$ decreases monotonically with increase in $M_m \beta_m$. Hence, $\Phi(P, W, r, m, l)$ decreases with increase in $m$ and can be lower and upper bounded as

$$\Phi(P, W, r, \infty, l) \leq \Phi(P, W, r, m, l) \leq \Phi(P, W, r, 1, l),$$

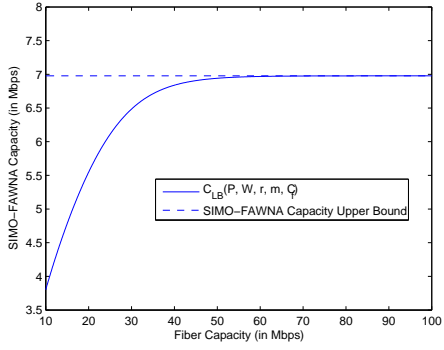

Fig. 2. Dependence of SIMO-FAWNA capacity on fiber capacity.

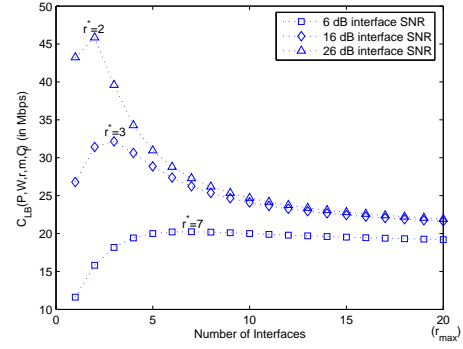

Fig. 3. Effect of the number of interfaces on $C_{\text{LB}}(P, W, r, m, C_f)$.

where, $\Phi(P, W, r, 1, l)$ and $\Phi(P, W, r, \infty, l)$ correspond to loss in capacity due to fixed rate, scalar and infinite dimensional vector quantization, respectively. Reduction in quantizer dimension reduces complexity at the interface but results in a capacity penalty. The maximum loss in capacity occurs when fixed rate, scalar quantizers are used at the wireless-optical interfaces.

### B. Effect of quantizer rate

We now study the effect of quantizer rate, $l$, on performance of the single-input, multiple-output FAWNA. Like the previous subsection, it suffices to consider the function, $\Phi(P, W, r, m, l)$. To make the expressions compact, define $\psi(l) \triangleq \frac{(1 - M_m \beta_m 2^{-l})}{1 + \frac{P M_m \beta_m 2^{-l}}{N_0 W}}$. Note that $\psi(l) = \Theta(2^{-l})$. Now,

$$\Phi(P, W, r, m, l) = W \log\left(1 + \frac{r[1-\psi(l)]\frac{P}{N_0 W}}{1 + r\psi(l)\frac{P}{N_0 W}}\right)$$
$$\leq \frac{P}{N_0} \cdot \frac{r[1-\psi(l)]\log(e)}{1 + r\psi(l)\frac{P}{N_0 W}} = O(2^{-l}).$$

and

$$\Phi(P, W, r, m, l) \geq \frac{P}{N_0} \cdot \frac{r[1-\psi(l)]\log(e)}{1 + r\psi(l)\frac{P}{N_0 W}}$$
$$- \frac{P^2 \log(e)}{2 N_0^2 W} \left[\frac{r[1-\psi(l)]}{1 + r\psi(l)\frac{P}{N_0 W}}\right]^2 = \Omega(2^{-l}).$$

To obtain these bounds, we use $x - \frac{1}{2}x^2 \leq \log_e(1+x) \leq x$. Hence,

$$\Phi(P, W, r, m, l) = \Theta(2^{-l}).$$

This implies that the capacity using the proposed scheme approaches the capacity upper bound (3), *exponentially* with quantizer rate. Also observe that $\Phi(P, W, r, m, \infty) = 0$. From (4), we see that the maximum value for the quantizer rate is dependent on the fiber capacity, number of interfaces and wireless bandwidth. Since $\Phi(P, W, r, m, l)$ decreases exponentially with $l$, it is minimum when $l$ is maximum, i.e.,

$$\min_{l \leq \frac{C_f}{rW}} \Phi(P, W, r, m, l) = \Phi(P, W, r, m, \frac{C_f}{rW}).$$

Let $C_{\text{LB}}(P, W, r, m, C_f)$ denote the lower bound to the capacity using the proposed scheme:

$$C_{\text{LB}}(P, W, r, m, C_f) \triangleq W \log\left(1 + \frac{rP}{N_0 W}\right)$$
$$- \Phi(P, W, r, m, \frac{C_f}{rW}).$$

Hence, the capacity for the proposed scheme approaches the SIMO-FAWNA capacity, exponentially with the capacity of the optical fiber. Note that though our scheme simply quantizes and forwards the wireless signals without source coding, it is near-optimal since the fiber capacity is much larger than the wireless capacity. This behavior is illustrated in figure 2, which is a plot of $C_{\text{LB}}(P, W, r, m, C_f)$ and the upper bound (3), versus fiber capacity. In the plot, we set $W = 1$ Mhz, $M_m \beta_m = 1$, $r = 5$ and $\frac{P}{N_0} = 25 \times 10^6$ sec$^{-1}$. Note that the fiber capacity required to achieve good performance is not large for an optical fiber, e.g. Gigabit Ethernet, OC-48, etc. which have speeds in the order of Gigabit/sec.

In the subsequent subsections, we analyze the effect of transmit power, number of antennas and wireless bandwidth on the capacity lower bound, $C_{\text{LB}}(P, W, r, m, C_f)$.

### C. Effect of transmit power on capacity lower bound

An increase in transmit power, $P$, leads to two competing effects. The first is increase in receive power at the interfaces, which increases wireless capacity. The second is increase in quantizer distortion, which reduces wireless capacity. The capacity lower bound,

$$C_{\text{LB}}(P, W, r, m, C_f) = W \log\left(1 + \frac{r(1 - M_m \beta_m 2^{-\frac{C_f}{rW}})\frac{P}{N_0 W}}{1 + \frac{P M_m \beta_m 2^{-\frac{C_f}{rW}}}{N_0 W}}\right),$$

increases monotonically with $\frac{r(1 - M_m \beta_m 2^{-\frac{C_f}{rW}})\frac{P}{N_0 W}}{1 + \frac{P M_m \beta_m 2^{-\frac{C_f}{rW}}}{N_0 W}}$, which in turn increases monotonically with $P$. Hence, the first effect always dominates and the capacity lower bound *increases* monotonically with transmit power.

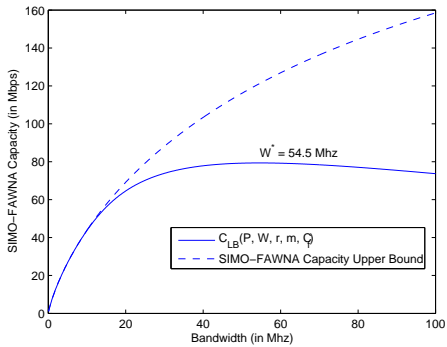

Fig. 4. Dependence of SIMO-FAWNA capacity on bandwidth.

*D. Effect of the number of interfaces on capacity lower bound*

Let us focus on the effect of the number of interfaces, $r$, on the capacity lower bound, $C_{\mathsf{LB}}(P,W,r,m,C_f)$. Since the quantization rate at the interface is never allowed to go below 1, the maximum number of interfaces possible is $r_{\mathsf{max}} = \lfloor \frac{C_f}{W} \rfloor$. Keeping all other variables fixed, the optimal number of interfaces, $r^*$, is given by

$$r^* = \arg\max_{r \in \{1,2,\ldots,r_{\mathsf{max}}\}} C_{\mathsf{LB}}(P,W,r,m,C_f).$$

For fixed wireless bandwidth and fiber capacity, an increase in the number of interfaces leads to two competing effects. First, wireless capacity increases due to receive power gain from the additional interfaces. Second, quantizer distortion increases due to additional interfaces sharing the same fiber, which results in capacity reduction. Hence, capacity doesn't increase monotonically with the number of antennas. Obtaining an analytical expression for $r^*$ is difficult. However, $r^*$ can easily be found by numerical techniques. Figure 3 is a plot of $C_{\mathsf{LB}}(P,W,r,m,C_f)$ versus $r$ for $W = 5$ Mhz, $M_m\beta_m = 1$, $C_f = 100$ Mbps. Plots are obtained for $\frac{P}{N_0} = 20 \times 10^6$ sec$^{-1}$, $200 \times 10^6$ sec$^{-1}$ and $2000 \times 10^6$ sec$^{-1}$. This corresponds to interface signal-to-noise ratios (SNR) of 6 dB, 16 dB and 20 dB, respectively. The corresponding values of $r^*$ are 7, 3 and 2, respectively. Observe that $r^*$ decreases with increase in interface SNR. This happens because when interface SNR is low, it becomes more important to gain power rather than to have fine quantization. On the other hand, when interface SNR is high, the latter is more important. Hence, as interface SNR decreases, $r^*$ tends towards $r_{\mathsf{max}}$.

*E. Effect of bandwidth on capacity lower bound*

We now analyze the effect of wireless bandwidth, $W$, on the capacity lower bound. Since the quantization rate is never allowed to go below 1, the maximum possible bandwidth is $\frac{C_f}{r}$. For fixed fiber capacity and number of interfaces, the optimal bandwidth of operation, $W^*$, is given by

$$W^* = \arg\max_{W \in [0, \frac{C_f}{r}]} C_{\mathsf{LB}}(P,W,r,m,C_f).$$

Since quantizer distortion as well as power efficiency increases with $W$, the behavior of the capacity lower bound with bandwidth is similar to that with the number of interfaces. Note that the quantization rate decays inversely with bandwidth. When the operating bandwidth is lowered from $W^*$, the capacity lower bound is lowered because the reduction in power efficiency is more than the reduction in quantizer distortion. On the other hand, when the operating bandwidth is increased from $W^*$, the loss in capacity from increased quantizer distortion is more than the capacity gain from increased power efficiency.

The optimal bandwidth, $W^*$, can be found by numerical techniques. Figure 4 shows the plot of the capacity lower bound and the upper bound (3) for $C_f = 200$ Mbps, $M_m\beta_m = 1$, $r = 2$ and $\frac{P}{N_0} = 100 \times 10^6$ sec$^{-1}$. The optimal bandwidth for this case is $W^* \sim 54.5$ Mhz.

V. CONCLUSIONS

In this work, we study a single-input, multiple-output FAWNA from a capacity view point. We propose a scheme and show that it has near-optimal performance when the fiber capacity is larger than the wireless capacity. We also show that for a given fiber capacity, there is an optimal operating wireless bandwidth and an optimal number of wireless-optical interfaces. The wireless-wireline interface has low complexity and does not require knowledge of the transmitter code book. The design also has extendability to FAWNAs with large number of transmitters and interfaces and, offers adaptability to variable rates, changing channel conditions and node positions. Future work may consider FAWNAs with multiple transmitters and examine the performance of various multiple access schemes.

ACKNOWLEDGEMENTS

The authors are thankful to Michelle Effros for helpful discussions. We also acknowledge the research grants NSF CNS-0434974, NSF 010727-005, Air Force award PY-1362 and DAWN 6897872.